\documentclass[figures]{epl}
\usepackage{bm}
\newcommand{\abag}{A_{{\rm Bag}}}
\newcommand{\tr}{\theta_{R}}
\newcommand{\eq}{\begin{equation}}
\newcommand{\en}{\end{equation}\noindent}

\title{Granular gravitational collapse and chute flow}

\author{D. Erta{\c s}\thanks{E-mail: \email{deniz.ertas@exxonmobil.com}} 
\and T. C. Halsey}

\shortauthor{D. Erta{\c s} \etal}

\institute{ExxonMobil Research and Engineering, 1545 Route 22 East, Annandale, NJ 08801,
USA}


\begin{document}
\maketitle

\begin{abstract}

Inelastic grains in a flow under gravitation tend to collapse into states in which the relative normal velocities of two neighboring grains is zero. If the time scale for this gravitational collapse is shorter than inverse strain rates in the flow, we propose that this collapse will lead to the formation of ``granular eddies", large scale condensed structures of particles moving coherently with one another. The scale of these eddies is determined by the gradient of the strain rate. Applying these concepts to chute flow of granular media, (gravitationally driven flow down inclined planes) we predict the existence of a bulk flow region whose rheology is determined only by flow density. This theory yields the experimental ``Pouliquen flow rule", correlating different chute flows; it also correctly accounts for the different flow regimes observed.

\end{abstract}

\section{Introduction}

Flows of hard granular systems are ubiquitous in nature and technology, yet are still poorly understood \cite{overall}. Granular systems typically have a twofold separation of energy scales: the typical energy of a particle is determined by gravity or some other body force (in a few instances by initial conditions), and is much larger than the thermal scale $k_B T$, yet much smaller than the scale required to appreciably deform the particle. 
Despite the smallness of $k_B T$ on the scale of granular energies, many treatments use a pseudo-temperature connected to the random part of the kinetic energy of a particle. Such treatments often link granular phenomena to the kinetic theory of gases. The ``granular gas" has an intrinsic rheology, and is driven by the external forcing.

\begin{figure} 
\twoimages[width=6cm]{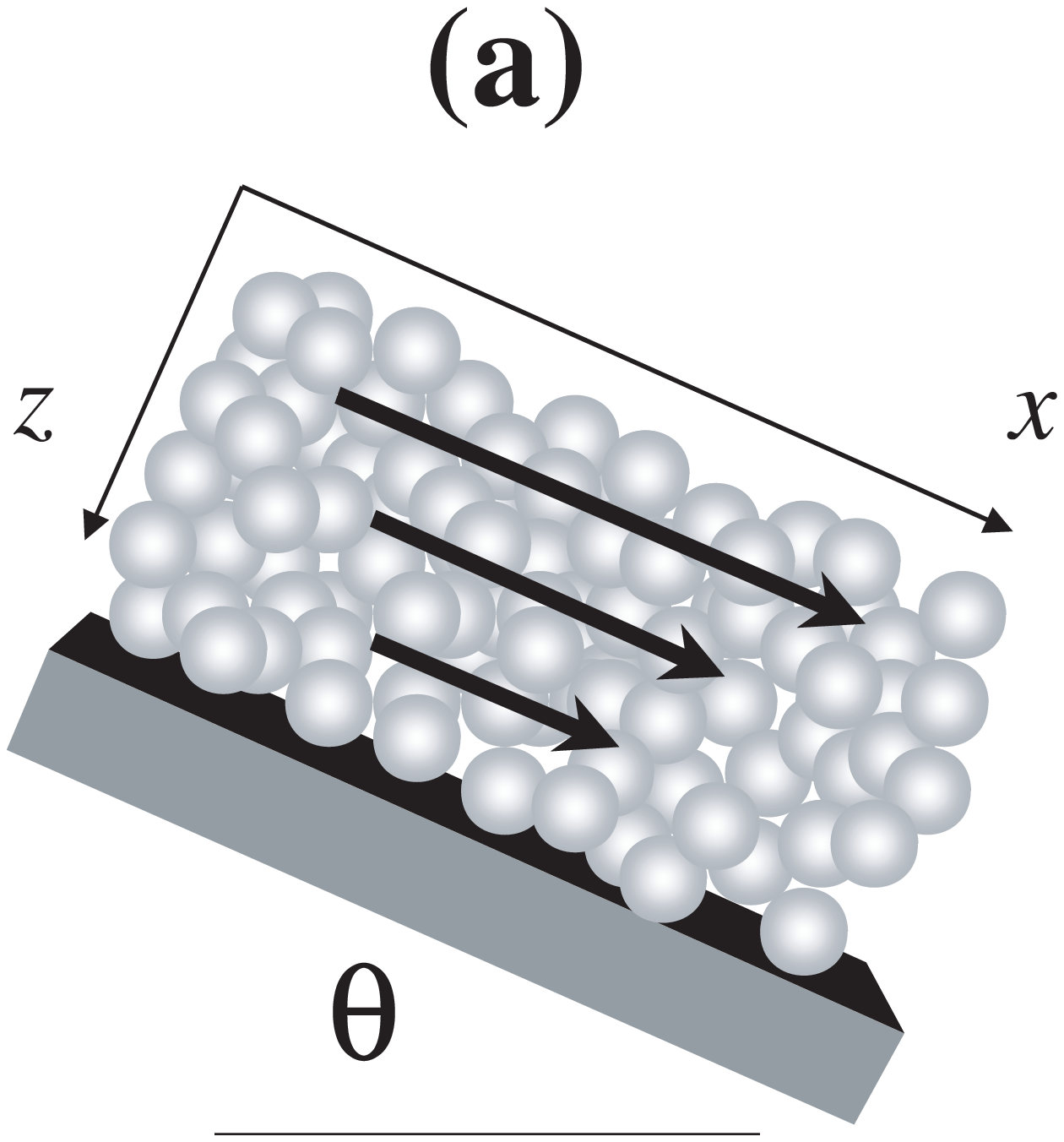}{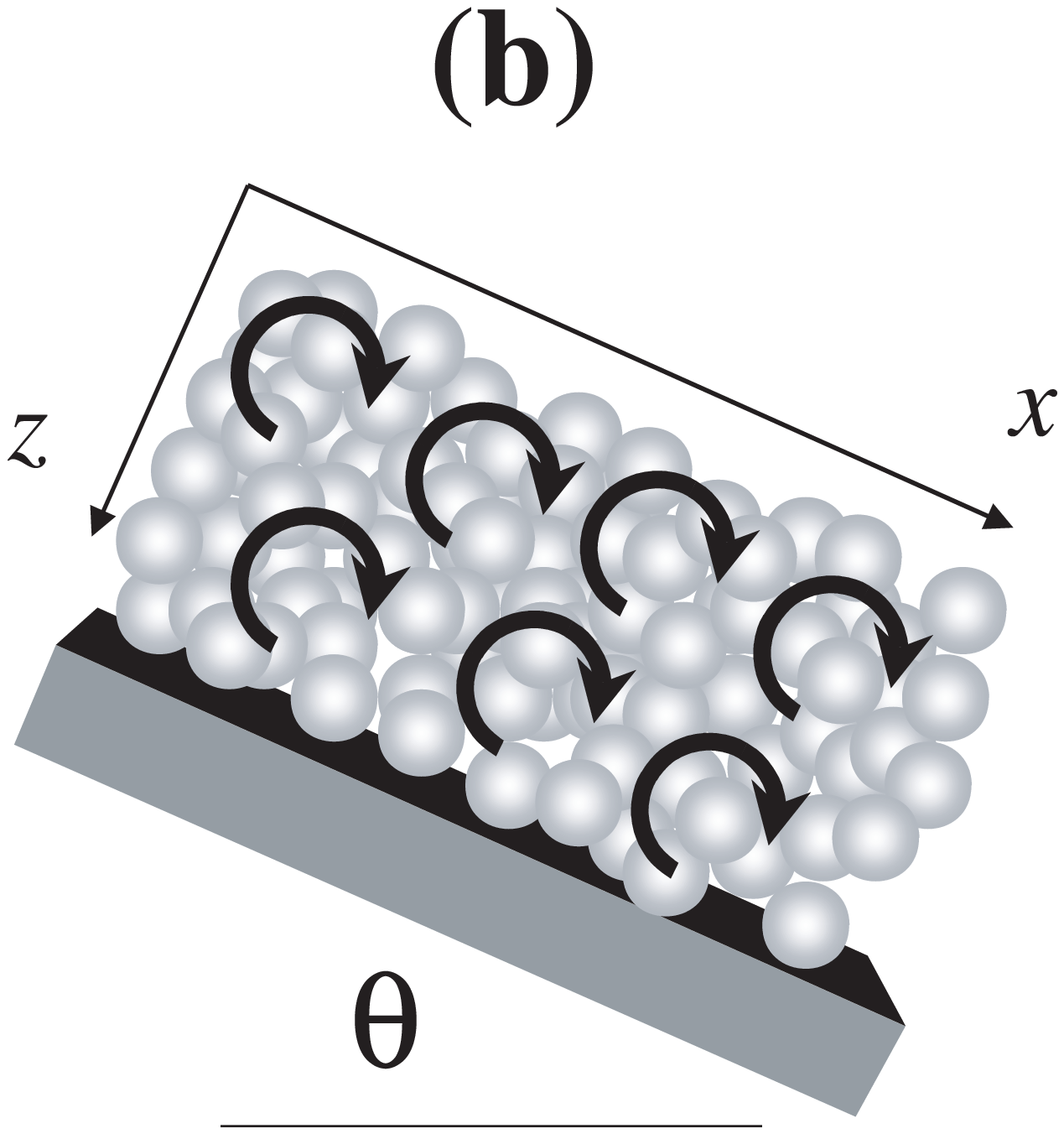}
\caption{(a) Chute flow is flow down a surface inclined at an angle $\theta$. The $x$-axis is chosen parallel to the flow, the $z$-axis perpendicular to the free surface. The $y$-axis is parallel to the vorticity of the flow, and is directed out of the page. (b) In the granular eddy picture, the motion of the particles is regarded as a superposition of the translation and rotation of granular eddies.} 
\label{fig1} 
\end{figure}

One of the pioneering treatments of this rheology was by Bagnold, who discussed chute flows, the gravitationally driven flow of a granular material down an inclined surface \cite{bagnold}. It is simplest to consider a flow of constant, fixed depth $H$, with the average velocity of the particles parallel to the free surface. The particles are spheres of monodispersed mass $M$ and radius $R$. We choose axes such that the direction of flow is $\hat x$, the direction perpendicular to the free surface of the flow is $\hat z$, and the direction parallel to vorticity is $\hat y$ (see Figure \ref{fig1}). The shear stress $\sigma_{xz}$ in such a flow is communicated by particles at slightly differing depths, whose velocities differ if $\partial_z v_x$ is non-zero. We expect the momentum transfer communicated by collisions between particles at different depths to be of the order of

\eq
\Delta p = M R \partial_z v_x.
\en

Furthermore, these collisions will occur at typical intervals of the order of

\eq
\Delta t = (\partial_z v_x)^{-1},
\en
from which we conclude that the typical collisional stress will be

\eq
\sigma_{xz}\sim \frac{1}{R^2}\frac{\Delta p}{\Delta t} = \frac{M}{R} 
(\partial_z v_x)^2.
\en

In a steady state flow down a surface inclined at an angle $\theta$, the $xz$ shear component of the stress tensor is determined by gravity to be

\eq
\sigma_{xz} = \rho g z \sin \theta,
\label{eq:stress}
\en
with $\rho$ the (local) mass density, which we here assume to be independent of $z$ (we will return to this point below.) We are measuring the depth of the pile $z$ from the free surface, at which $z=0$. 

If $\sigma_{xz} \propto z \propto (\partial_z v_x)^2$, then since $v_x=0$ at the base $z=H$ if the boundary pins the flow, we immediately obtain

\eq
\partial_z v_x = - A_{{\rm Bag}} \sqrt{z},
\en
defining the coefficient $\abag$, or 

\eq
v_x(z) = \frac{2 \abag }{3}(H^{3/2} - z^{3/2}).
\en

While there have been a variety of authoritative experimental studies of chute flow \cite{savage}, as well as intricate theoretical discussions of the rheologies to be expected on general grounds \cite{theory}, we have been particularly inspired by the recent work of Pouliquen \cite{pouliquen}. Pouliquen studied the behavior of chute flows as a function of inclination angle $\theta$ and height of the flow $H$. He found that for small values of $\theta$ or height $H$, no flow took place. 

With the increase of either $\theta$ or $H$ such that an angle of repose line $\theta_{{\cal R}} (H)$ was passed, a region of steady-state flow was entered. Finally, for values of $\theta$ above a maximum $\theta_M$, the flows continuously accelerated, and no steady-state flow was observed.

The dominant observational fact about the steady-state flows is the 
``Pouliquen flow rule," which connects the vertically averaged velocity $u$ of a flow of height $H$ with the height $H_{stop}$ at which flow ceases for a chute of that inclination $\theta$. [The angle of repose $\theta_{{\cal R}} (H)$ is the inverse of the function $H_{stop} (\theta)$.] The Pouliquen flow rule gives a scaling form for the 
average velocity $u$ of the flow,

\eq
u \equiv \frac{1}{H} \int_0^H v_x (z) dz,
\en
of
\eq
\frac{ u }{\sqrt{gH}} = \beta \frac{H}{H_{stop}},
\label{eq:rule}
\en
and accounts well for experimental data with $\beta=0.136$.

The scaling $u \propto H^{3/2}$ in the Pouliquen flow rule is consistent with the Bagnold rheology. But the Pouliquen flow rule also connects the coefficient $\abag (\theta)$ with {\it the thickness of the pile at that inclination below which flow arrests}, which would not be expected from the Bagnold point of view. 

Note that we would have obtained dimensionally the Bagnold result for the rheology had we claimed that the stress should obey
\eq
\sigma_{xz} = \mu \partial_z v_x \sim \rho R^2 (\partial_z v_x)^2,
\label{eq:viscosity}
\en
where we have made the substitution for the viscosity $\mu \sim \rho R^2 
\partial_z v_x$ on grounds that a granular flow has no other obvious local length or time scales than $R$ for the length scale and $(\partial_z v_x)^{-1}$ for the time scale. The heart of the Bagnold approach thus lies in the assumption that these are in fact the only local scales.  Note that the gravitational constant $g$ does not figure directly in either of these scales. However, the Pouliquen flow rule implies that this rheology does depend both upon $g$ and upon the thickness of the arresting pile $H_{stop}$, which is hardly local information. Thus the Pouliquen flow rule appears to be inconsistent with any assumption of a purely local rheology comparable to that of a granular gas \cite{bocquet}. This, and other considerations, have motivated some authors to build non-local models for the rheology \cite{nonlocal}.

The broad features of Pouliquen's conclusions have been confirmed by a series of numerical studies in which these authors have participated \cite{grest}. For relatively thin piles, the Bagnold rheology breaks down (as also seen in experiment, \cite{surface}), but the thicker piles show a Bagnold rheology and obey the Pouliquen flow rule, albeit with a slightly larger value of $\beta$ (The crossover is examined numerically in \cite{layer}).  However, the assumption of the Bagnold or granular gas approach that the stress is mostly transmitted through collisions seems not to be true in these numerical studies; stress seems instead to be transmitted primarily through relatively long-lived contacts between particles. The density in the interior of the piles is independent of depth, consistent with the assertion made in Eq.(\ref{eq:stress}).

In this treatment, we eschew granular gas approaches, and we do not assume the existence of any rheology independent of the gravitational character of the flow. We show that gravitation combined with particle inelasticity is able to dissipate a significant fraction of the system's kinetic energy over time scales short compared to the inverse strain rate. Given this fact, it is natural to posit the existence of ``granular eddies," gravitationally collapsed networks of particles, which move coherently and whose properties determine the flow rheology. The rheology that follows from this picture agrees with the Pouliquen flow rule, and also gives a simple explanation for the different flow regimes observed by Pouliquen.

In this report, we first analyze the phenomenon of gravitational collapse for inelastic particles. We then introduce the granular eddy picture, and relate the eddy size to flow properties. We specialize to the case of chute flows, determining the rheology, and accounting for the principal features of the observed phenomenology of these flows. 

\section{Gravitational collapse}

Consider an inelastic ball with a coefficient of restitution of $\epsilon$, bouncing on a rigid horizontal surface. It is elementary to show that if its normal velocity at first impact is $v_0$, then after a finite time $\tau_{gc}$ it will come to rest, with

\eq
\tau_{gc} = \frac{2 v_0}{g} \frac{\epsilon}{1-\epsilon}.
\en
(A similar result obtains if we take a more realistic ball with a Hertzian contact force and a visco-elastic dissipation; here we restrict ourselves to the simplest case.)

Now consider a particle in a granular flow. Suppose that $\tau_{gc}$ is short compared to the time scales for its neighboring particles to rearrange themselves. This latter time scale is the inverse of the local shear rate $\dot \gamma$; for chute flow $\dot \gamma = \partial_z v_x$. In this short ``collapse time" regime, an individual particle will rapidly come to rest upon the particle or particles beneath it in the flow; it may then roll about, possibly arriving in a stable configuration with three particles beneath it. However, its motion is strongly constrained by its gravitational collapse, and its motion is strongly correlated with that of the particles with which it is in contact.

Thus we can envision large aggregations of particles coming into existence,each of whose motions with respect to its neighbors is at most of a rolling kind. Influenced by pictures of turbulence as a superposition of eddies, we term these aggregations ``granular eddies" (see Fig.~\ref{fig1}).

\section{Granular eddy size} 

Consider an eddy of scale $\ell$, whose center of mass is at a position $z_0$. The local average velocity can be expanded as

\eq
v_x(z) = v_x(z_0) + (z-z_0) \partial_z v_x (z_0) + \frac{1}{2} 
(z-z_0)^2 \partial_{zz} v_x (z_0) + \cdots
\en

While the first and second terms can be matched by an eddy whose center of mass moves at a velocity $v_x (z_0)$ and which rotates at an angular velocity $\omega = \partial_z v_x (z_0)$, the third term in this series is incompatible with the rigid-body rotation of an eddy. We write this incompatible velocity at the eddy boundary as

\eq
v_{ic} = \frac{1}{2} \ell^2 \partial_{zz} v_x. 
\en

Supposing this incompatibility is relieved by internal strains of the eddy on the scale $\ell$ of the eddy itself, the time scale corresponding to variation of these ``incompatibility strains" is 

\eq
\tau_{ic} \equiv \frac{\ell}{ v_{ic}} = \left (\frac{\ell}{2} 
\partial_{zz} v_x \right )^{-1}.
\en
On this time scale the environment of particles at the boundary of an eddy will inevitably change as that eddy conforms with the surrounding flow.

Since the radius $\ell$ determines the location of the outer perimeter of the eddy, then we anticipate that particles associated with neighboring eddies will typically be colliding with the eddy with relative velocities comparable to $v_c = 2 \omega \ell$. The characteristic gravitational collapse time associated with these collisions will be

\eq
\tau_{gc}=\frac{2 v_c}{g} \frac{\epsilon}{1-\epsilon}
\en
so that the criterion for the particles at the eddy surface to be able to gravitationally collapse before their environment is altered by incompatibility strains is

\eq
\frac{\tau_{gc}}{\tau_{ic}} = \left( \frac{4 \ell \partial_z v_x}{g} 
\frac{\epsilon}{1-\epsilon}\right) \left( \frac{\ell}{2} 
\partial_{zz} v_x \right) < {\tilde a},
\en
where $\tilde a$ is an unknown numerical constant of $O(1)$. The maximum value of $\ell$ consistent with this relation is determined by

\eq
\ell_e^2 \left[ \partial_z (\partial_z v_x)^2\right] = {\tilde a}g 
\frac{1-\epsilon}{\epsilon}.
\label{eq:eddyell}
\en

We expect that this maximum value will set the scale of the eddies, since eddies smaller than this size will tend to grow as more and more particles collapse onto their perimeters, and eddies larger than this sizewill lose particles from their perimeters.

\section{Phenomenology of chute flow}

Let us use dimensional analysis to define an effective ``viscosity length scale" $\ell_\nu$ by inverting the Bagnold scaling relation given by Eq.(\ref{eq:viscosity}) after substituting this new length scale instead of $R$. This yields

\eq
\sigma_{xz}\equiv\rho\ell_\nu^2(\partial_z v_x)^2.
\en
For chute flow, this can be combined with Eq.(\ref{eq:stress}) to give

\eq
\ell_{\nu} = \sqrt {\frac{g z \sin \theta }{(\partial_z v_x)^2}}.
\label{eq:ellmu}
\en

We now make a different scaling assumption than that of Bagnold, which is that the length scale appearing in Bagnold's argument should be set by the eddy size $\ell_e$ instead of the particle size $R$, i.e.,

\eq
\ell_\nu^2=\ell_e^2\left(1+\tilde b \frac{R}{\ell_e} + \cdots \right),
\label{eq:ellr} 
\en
where the unknown numerical constant $\tilde b$ accounts for the leading order finite-size corrections due to the existence of the particle size $R$. In other words, we are assuming that (with the exception of these corrections in $R$), this length scale is the unique length scale determining the bulk rheology of the granular dispersion. Then $\ell_e$ and $\partial_z v_x$ are jointly determined by simultaneous solution of Eqs.~(\ref{eq:eddyell}),(\ref{eq:ellmu}) and (\ref{eq:ellr}).

If the angle of inclination is too small, there is no solution, in particular, for

\eq
\theta < \tr\equiv\sin^{-1}\left( \tilde a \frac{1-\epsilon}{\epsilon}\right).
\en
For $\theta>\tr$, we have

\eq
\label{eq:eddysize}
\tilde b \frac{R}{\ell_e} = \frac{\sin \theta}{\sin \tr}-1,
\en
which fixes the eddy size $\ell_e$. Finally, we find

\eq
\partial_z v_x = -\abag \sqrt{z}
\en
with

\eq
\abag= \frac{\sqrt{g \sin \tr}}{\ell_e } = \frac{\sqrt{g \sin \tr}}{\tilde b R} 
\left(\frac{\sin \theta}{\sin \tr}-1 \right).
\en

We have up to now ignored the question of the density of the flow. Let us suppose that the eddies themselves have a fixed density $\rho_0$, independent of their size. Then the medium as a whole can have a density that differs from this (``free volume") only due to a presumably lower density in the regions, of typical scale $R$, that separate different eddies from one another. We conclude that

\eq
\rho = \rho_0 \left( 1 - \tilde c\frac{R}{\ell_e} + \cdots \right)
\en
or

\eq
\frac{\rho_0-\rho}{\rho_0} = \frac{\tilde c}{\tilde b} 
\left(\frac{\sin \theta}{\sin \tr}-1\right),
\en
where $\tilde c$ is yet another unknown numerical constant. Note that for chute flow, the eddy size $\ell_e$ given by Eq.(\ref{eq:eddysize})is independent of depth $(z)$, consistent with the assumption of a depth-independent density $\rho$. 

Of course, the eddy size $\ell_e$ is not entirely unrestricted. When the eddies get too large to be accomodated within the height $H$ of the flow , i.e., for $\ell_e \sim H/2$, we expect flow arrest to occur since all particles in the flow become connected to the bottom surface through the contact network. This gives the thinnest flowing pile at a given angle $\theta$ in terms of a new constant $\tilde d \sim 1$, 

\eq
H_{stop}(\theta)=\tilde d \ell_e(\theta)=\tilde b \tilde d R
\left(\frac{\sin\theta}{\sin\tr}-1\right)^{-1},
\en
or, equivalently, the lowest possible angle of stable flow at a given pile height $H$,

\eq
\theta_{{\cal R}} (H) = \sin^{-1}\left(
\sin\tr\bm{\left(} 1 + \tilde b \tilde d \frac{R}{H}\right)
\bm{\right)}.
\en

On the other hand, gravitational collapse ceases to stabilize the flow when $\ell_e\sim R$, corresponding to an upper limit of stability

\eq
\theta_M = \sin^{-1}\bm{(}\tilde e \sin (\tr)\bm{)},
\en
which is independent of the flow height $H$. Here $\tilde e \sim 1$ is a further unknown proportionality constant. Note that we can now re-interpret $\tr$ as the limiting value of the angle of repose as the thickness of the flow $H \to \infty$.

\begin{figure}
\oneimage[width=6cm]{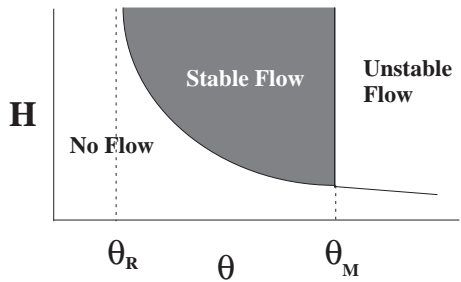}
\caption{The granular eddy picture predicts that the angle of repose line separating the region of no flow from that of stable flow depends on $H$. The angle of repose line approaches a fixed angle $\tr$ as $H \to \infty$. The maximum angle of stable flow $\theta_M$ does not depend on $H$ in this picture, except at small values of $H$ where the stable flow regime disappears.}  
\label{fig2}
\end{figure}

Since this is a scaling theory, it is not possible to make a quantitative comparison between these predictions and numerical results such as those of Ref.~\cite{grest}. However, for $\theta-\tr \ll 1$, the dependences on tilt angle for a thick flow ($H \gg R$) such as $\partial_z v_x = -\abag \sqrt{z}$ with $\abag \propto (\theta-\theta_R)$, and $\rho_0 - \rho \propto (\theta-\theta_R)$ are borne out by these numerical results. Also, the Pouliquen flow rule Eq.(\ref{eq:rule}) is recovered with

\eq
\beta = \frac{2\tilde d}{5}\sqrt{\sin \tr}.
\en
Finally, the form of the phase boundaries (see Fig.~\ref{fig2}) is in agreement with both numerical and experimental results.


In this Letter we have addressed the form of the bulk rheology for chute flows; our conclusions regarding this rheology should hold in portions of the pile for which the computed eddy scale $\ell$ is less than the distance to the boundaries. Clearly there will be both upper and lower boundary layers in which this is impossible. We have not addressed the structure of these boundary layers. Although chute flow does seem to have a bulk-rheology dominated regime, this may not be the case with all flow geometries. In some flows the structure of the boundary layers may dominate in determining the characteristics of the flow.

\acknowledgments

We are grateful to P.M. Chaikin, J. Landry, and L. Silbert for helpful discussions.

\end{document}